\documentclass[showpacs,twocolumn,prl,superscriptaddress]{revtex4}

\usepackage{amsmath}
\usepackage{amssymb}
\usepackage{graphicx}
\usepackage{bm}

\begin{document}

\title{Magneto-optical control of light collapse in bulk Kerr media}
\author{Y. Linzon$^{*}$}
\affiliation{Universit\'{e} du Quebec, Institute National de la
Recherche Scientifique, Varennes, Quebec J3X 1S2, Canada}
\author{K. A. Rutkowska}
\affiliation{Universit\'{e} du Quebec, Institute National de la Recherche Scientifique,
Varennes, Quebec J3X 1S2, Canada}
\affiliation{Faculty of Physics, Warsaw University of Technology, Warsaw PL-00662, Poland}
\author{B. A. Malomed}
\affiliation{Department of Physical Electronics, School of Electrical Engineering,
Faculty of Engineering, Tel Aviv University, Tel Aviv 69978, Israel}
\author{R. Morandotti}
\affiliation{Universit\'{e} du Quebec, Institute National de la Recherche Scientifique,
Varennes, Quebec J3X 1S2, Canada}
\date{\today}

\begin{abstract}
Magneto-optical crystals allow an efficient control of the
birefringence of light via the Cotton-Mouton and Faraday effects.
These effects enable a unique combination of adjustable linear and
circular birefringence, which, in turn, can affect the propagation
of light in nonlinear Kerr media. We show numerically that the
combined birefringences can accelerate, delay, or arrest the
nonlinear collapse of (2+1)D beams, and report an experimental
observation of the acceleration of the onset of collapse in a bulk
Yttrium Iron Garnet (YIG) crystal in an external magnetic field.
\end{abstract}

 \maketitle

Waves propagating in multidimensional self-focusing media are
subject to instabilities that lead to the catastrophic collapse
after a finite propagation distance, followed by beam filamentation
or material damage
\cite{Rev1,Rev2,Hydro,matterwave,BECcontrol,Fibich1,Fibich23,Fibich4,control12,control3,control4}.
In the (2+1)D [(2+1)-dimensional)] setting, above a certain
threshold value of the input power, the critical collapse driven by
focusing nonlinearities is a universal scenario, observed in
high-power excitations of plasmas \cite{Rev1,Rev2}, hydrodynamical
systems \cite{Hydro}, Bose-Einstein condensates (BECs)
\cite{matterwave,BECcontrol}, and optical media
\cite{Rev2,Fibich1,Fibich23}. In particular, in optical pulse
propagation through amorphous media and crystals without special
symmetries, the dominant nonlinearity is the Kerr (cubic) effect,
which is modeled by the nonlinear Schr\"{o}dinger equation (NLSE)\
\cite{Boyd}. Collapsing beams in Kerr media were studied in detail,
especially in the course of the past decade
\cite{Fibich1,Fibich23,Fibich4,control12,control3,control4}.

The challenge to \emph{control} the wave collapse (and in particular
to mitigate its detrimental effects) has recently drawn much
attention \cite{control12,BECcontrol,control3,control4}. As recently
demonstrated, the collapse distance of ultra-intense laser pulses in
air can be controlled by passive optical elements \cite{control12},
and in BECs the collapse time is strongly affected by the so-called
Feshbach-resonance technique \cite{BECcontrol}. In condensed optical
media, where the collapse occurs with pulse energies far below the
creation of plasma \cite{Fibich1,Fibich23}, a recently proposed
scheme for collapse management relies on the use of a layered
structure with the nonlinearity strength alternating in the
longitudinal direction \cite{control3} (a similar mechanism was
proposed for the stabilization of BECs in the 2D case
\cite{control4}). However, this scheme is difficult to implement in
bulk media, and such structures may give rise to linear losses
induced by the reflection of light from interfaces between the
alternating layers.

An alternative approach to control the transition to the collapse
may be provided by optical birefringence, which promotes energy and
phase transfer between the polarization components of the beam. The
interplay between birefringence and nonlinearity is known to induce
coupling of the polarization rotation and a characteristic temporal
evolution of solitons in optical fibers \cite{birefringence}. In
this paper, we explore the birefringence as a tool for the
\textquotedblleft management" of the collapse of (2+1)D beams, which
may be relatively easily implemented in affordable experimental
conditions by using magneto-optical (MO) effects
\cite{MO1,CM,Faraday-YIG,isolator}, while avoiding reflection
losses. To this end, we present numerical and experimental studies
of the combined effects of linear and circular birefringences on the
dynamics of collapsing beams in a bulk self-focusing Kerr medium. We
find that the onset of collapse can be accelerated, delayed, or even
suppressed, at certain values of the combined birefringence
strengths. We also show that the required linear and circular
birefringence can be induced in a transparent MO Yttrium Iron Garnet
(YIG, Y$_{3}$Al$_{5}$O$_{12}$) crystal by the application of an
external dc magnetic field, thus opening up the perspective of using
MO effects to generate and control various nonlinear phenomena in
optics. In our experiments, an adjustable balance of linear and
circular birefringences was realized via a combination \cite{MO1} of
the Cotton-Mouton (or Voigt) \cite{CM} and Faraday
\cite{Faraday-YIG} MO effects in a bulk YIG crystal. Following the
propagation of femtosecond pulses in the crystal, we observed a
controllable decrease of the threshold input power necessary for the
onset of collapse at the output facet as a function of the magnetic
field. This also constitutes a first experimental study in nonlinear
optics where birefringence can be switched on and varied
continuously in an adjustable fashion.

The evolution of the complex electric-field amplitudes, $u_{r}$ and
$u_{l}$, representing the right- and left-circular polarizations
(RCP and LCP), in the presence of a Kerr nonlinearity and combined
linear and circular birefringences, obeys the coupled NLSEs, in the
scaled form \cite {birefringence}:
\begin{equation}
\begin{array}{c}
i\frac{\partial u_{r}}{\partial z}+\frac{1}{2}\nabla _{\bot
}^{2}u_{r}+bu_{r}+cu_{l}+(|u_{r}|^{2}+2|u_{l}|^{2})u_{r}=0, \\
i\frac{\partial u_{l}}{\partial z}+\frac{1}{2}\nabla _{\bot
}^{2}u_{l}-bu_{l}+cu_{r}+(|u_{l}|^{2}+2|u_{r}|^{2})u_{l}=0
\end{array}
\label{eqs}
\end{equation}

\noindent where $z$ is the propagation axis, $\nabla _{\bot }^{2}$
is the transverse Laplacian, while $b$ and $c$ are the strengths of
the circular and linear birefringences, respectively. As evident
from Eqs. (\ref{eqs}) the linear and circular birefringences account
for, respectively, the rates of linear amplitude mixing and phase
shift between the RCP and LCP fields. In terms of the RCP and LCP,
the ratio between the cross- and self-phase modulation coefficients
(XPM/SPM) is $2$ for cubically-symmetric crystals, including YIG
\cite{Boyd,birefringence}.

Assuming vorticity-free solutions with circular symmetry, (2+1)D
Eqs. (\ref{eqs}) reduce to a (1+1)D form, in terms of $z$ and the
radial coordinate $R$. We consider input Gaussian beams,
$u_{r}(R,z=0)=A\exp\left(-\frac{R^{2}}{2\rho^{2}}\right)\cos\theta$
and
$u_{l}(R,z=0)=A\exp\left(-\frac{R^{2}}{2\rho^{2}}\right)\sin\theta$,
with normalized input width $\rho$=1, amplitude $A$, and symmetric
polarization content $\theta$=$\pi/4$. This choice corresponds to an
unchirped Gaussian profile launched with an horizontal linear
polarization. Since the input does not carry vorticity, the
circular-symmetric structure of the solutions is not subject to an
azimuthal modulational instability \cite{Fibich23}.

\begin{figure}[b]
\includegraphics[width=8cm]{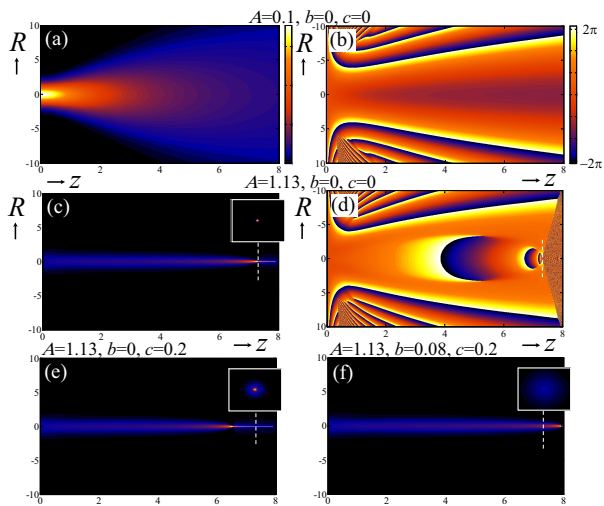}
\caption{ (Color online) Propagation maps of $u_{r}(R,z)$ obtained
from direct solutions of Eq. (\protect\ref{eqs}). (a)-(d): Zero
birefringences, with (a),(b) low and (c),(d) high input powers. The
pairs of panels (a),(c) and (b),(d) display the evolution of the
intensity, $|u_{r}(R,z)|^{2}$, and phase of $u_{r}(R,z)$,
respectively. (e), (f): Intensity maps in the presence of the
birefringences. Numerical parameters are indicated above each
respective panel.}
\end{figure}

\begin{figure}[t]
\includegraphics[width=8cm]{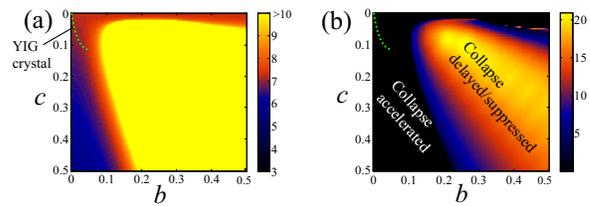}
\caption{ (Color online) Evolution maps in the parameter space
$(b,c)$ (see the text), as obtained from direct simulations of Eqs.
(\ref{eqs}), for $A$=$1.13$ and 0$\leq$$z$$\leq$50, up to $z=50$ or
the value $z_{\mathrm{coll}}<50$ in which the collapse occurs. The
solutions that do not collapse up to $z=50$ are represented in
yellow areas. The dashed (green) curves show the calculated
dependence between $b$ and $c$ corresponding to crystalline YIG in
an external magnetic field. (a) $z_{\mathrm{coll}}$, and (b) beam
width at $z_{\mathrm{coll}}$.}
\end{figure}

Figure 1 shows propagation maps obtained by direct simulations of
Eqs. (\ref{eqs}). In the absence of birefringence [Figs. 1(a)-(d)],
the RCP and LCP components are coupled only via the XPM term, which
becomes significant only close to the collapse point. While a low
input power excitation [Figs. 1(a,b)] results in beam diffraction,
the collapse occurs after a finite propagation distance [for
instance, $z=7.3$ in Fig. 1(c)] when the input power exceeds the
critical level \cite{Fibich4}. A substantial converging portion of
the phase fronts emerges near the collapse point, see Fig. 1(d). The
collapse can be \emph{accelerated} by the introduction of linear
birefringence, as shown in Fig. 1(e). Circular birefringence, if
acting alone, does not affect the collapse, since the respective
terms in Eqs. (\ref{eqs}) can be eliminated by a straightforward
transformation. When both of the birefringences are present, the
propagation distance necessary for the onset of collapse can be
\emph{extended}, i.e., the onset of the collapse is delayed [Fig.
1(f)] due to the interplay between amplitude and phase mixing. For
low $b$ and $c$ values, such as those used in Fig. 1(e,f) and
typically achievable in experiment, the differences in the evolution
of $u_{r}$ and $u_{l}$ are marginal, i.e., the RCP and LCP beam
components feature the same collapse dynamics. With larger
birefringence parameters the differences between the components
become substantial; however, such large birefringence values were
not accessible in the current experiment, see below.

The results of systematic simulations are summarized in Fig. 2 by
means of maps in the plane of $(b,c)$. Panels (a) and (b) show,
respectively, the values of $z_{\mathrm{coll}}$ where the beam
collapses, if $ z_{\mathrm{coll}}<50$, and the final values of the
beam width, $\rho (z=50)$; in cases where the collapse occurs at
$z_{\mathrm{coll}}<50$, the final width is set as $\rho =0$ [black
areas in (b)]. While Figs. 1 and 2 display the results for the RCP
component, the LCP maps are similar in the entire range considered.
As seen in Fig. 2, for a given input power the domination of
amplitude mixing between RCP and LCP ($c>b$) usually results in an
acceleration of the collapse, while dominant phase mixing ($b>c$)
can lead to the delay or effective suppression of the collapse.

In the experiment, a bulk YIG single-crystal was placed with the
easy crystallographic axis $[100]$ parallel to $z$, cf. Ref.
\cite{isolator}. These cubic dielectric crystals are highly
transparent for optical signals in the near-infrared and exhibit
large MO transmission coefficients, owing to their ferrimagnetic
phase \cite{CM,Faraday-YIG}. We chose to work at the wavelength of
$1.2$ $\mathrm{\mu}$m, which offers an optimal trade-off between the
magnitudes of the MO coefficients and the absorption losses
\cite{Faraday-YIG}. The temporal dispersion in YIG is normal and
weak in the near-infrared \cite{YIG-param}, and hence a high power
beam is free from the development of temporal modulational
instabilitiess \cite{Fibich23}.

The application of an external magnetic field $\mathbf{H}$
perpendicular to the propagation axis renders the crystal optically
uniaxial, with the optical axis parallel to $\mathbf{H}$ and the
respective linear birefringence proportional to $H$ \cite{MO1,CM}.
The largest phase retardation associated with the linear
birefringence, for components of the wavevector $\mathbf{k}$
perpendicular to the magnetic field, is $0.45$ \textrm{$\mu $}m/cm
in YIG \cite{CM} at a saturation field of $H=800$ G. This
corresponds to a saturation value of $c=0.1$ in terms of Eq. (1) for
our crystal. For wavevector components parallel to the magnetic
field, the Faraday effect induces different refractive indices for
the RCP and LCP waves (i.e., circular birefringence), with a
respective polarization phase retardation of $2$ $\mathrm{\mu }$m/cm
at the saturation level \cite{Faraday-YIG}. While the Faraday effect
is predominant in the $\mathbf{k}\parallel \mathbf{H}$ geometry
\cite{isolator}, it is weaker in the present setting, as only near
the collapse point the wavevectors feature significant lateral
components. An estimation corresponding to the case shown in Figs.
1(c),(d) yields an effective saturation value of $b=0.05$. Below
saturation, the Faraday coefficient is proportional to the magnetic
field, $b\sim H$, while the Cotton-Mouton coefficient depends on $H$
quadratically, i.e. $c\sim H^{2}$ \cite{MO1,CM,Faraday-YIG},
implying a parabolic dependence between the birefringence
coefficients in a given external magnetic field, $b\sim c^{2}$.
Calculated relations between $b$ and $c$ for the YIG crystal at
different magnetic fields are shown by the dashed curves in Fig. 2.

\begin{figure}[t]
\includegraphics[width=8cm]{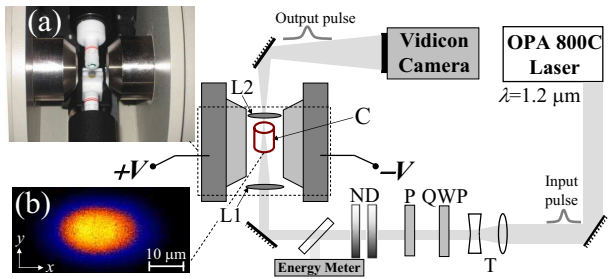}
\caption{ (Color online) Experimental setup schematics. OPA, optical
parametric amplifier laser; T, cylindrical telescope; QWP, quarter
wave plate; P, polarizer; ND, neutral density attenuators; L1,L1,
aspheric coupling and imaging lenses; C, crystal sample. Insets: (a)
Photograph of the magnet's interior with lenses L1, L2 and crystal
C; (b) The beam's waist profile at the input facet of the crystal.}
\end{figure}

The experimental setup is sketched in Fig. 3. We used a Spectra
Physics model 800C Optical Parametric Amplifier laser system,
delivering pulses of $200$ fs duration at a repetition rate of $1$
KHz with peak powers $\leq 50$ MW. The input beam was shaped by
means of a cylindrical telescope (T), followed by the combination of
a quarter wave plate (QWP) and a polarizer (P), which were used to
fix the input linear polarization parallel to $\mathbf{H}$ (i.e.
horizontal). Variable neutral density filters (ND) were used to
control the input power. The YIG sample was placed inside an
electromagnet (GMW model 3470). The application of a driving voltage
between the poles ($\pm V$) induced a uniform out-of-plane dc
magnetic field in the crystal. A digital Tesla-meter (Group3
DTM-133) was used to calibrate the free-space magnetic field and
verify its homogeneity in the sample. A combination of aspheric
lenses, L1 and L2, was used for coupling and imaging, respectively.
Importantly, the lenses and crystal were mounted on nonmagnetic
pyrex holders, see Fig. 3(a). Standard metallic mounts, which
usually hold microscope objective lenses, were not used, as we found
that they strongly affect the uniformity of the magnetic field
within the sample. The sample input facet was placed at the focal
plane of L1. The beam's profile at this plane is shown in Fig. 3(b).
The FWHMs of this elliptical input beam were $30$ and $15$
$\mathrm{\mu }$m along the major (\emph{x}) and minor (\emph{y})
axes. The beam profile at the sample output facet was imaged by L2
onto a Vidicon infrared camera.

\begin{figure}[t]
\includegraphics[width=8cm]{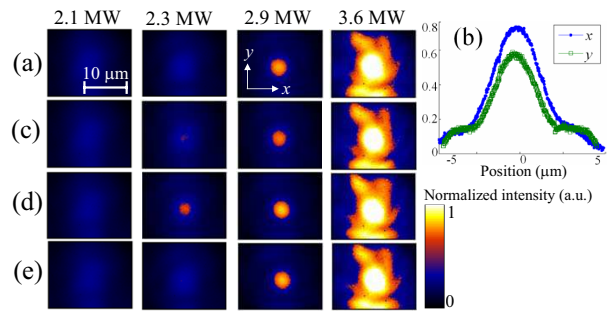}
\caption{ (Color online) (a) Output beam profiles as a function of
the input peak power, in the absence of an external magnetic field.
(b) Line-outs along $x$ (full circles) and $y$ (hollow rectangles)
around the central spot of the collapsed ($2.9$ MW) output beam.
(c)-(e) Output beam images, as function of the input power
(horizontal) and the external magnetic field (vertical), for the
fields: $H=200$ G (c), $400$ G (d), $800$ G (e).}
\end{figure}

Experimental results are summarized in Fig. 4. Images of the output
beam as a function of the input peak power, in the absence of a
magnetic field, are shown in Fig. 4(a). At the critical power of
$2.9$ MW, a central high-power spot appears in the diffracting
background. This spot is circularly symmetric [see Fig. 4(b)],
although the input beam profile was elliptical, with the
above-mentioned major-to-minor axes ratio of $2:1$. The spot fits a
Townes profile \cite{Fibich1}, which indicates that it is generated
by the onset of the collapse. The corresponding critical power
complies with the theoretical prediction for collapse
\cite{Fibich4}, using $\lambda_{0}=1.2$ $\mathrm{\mu}$m, the linear
refractive index of YIG $n_{0}=1.83$ and its Kerr coefficient
$n_{2}=7.2\times 10^{-16}$ cm$^{2}$/W \cite{YIG-param}. The
corresponding collapse distance prediction with $A=1.13$,
$z_{\mathrm{coll}}=7.3$ [Figs. 1(c),(d)], when rescaled back into
physical units using the well-known transformations
\cite{Fibich1,Fibich23,Fibich4} and assuming a circular input beam
of 10 $\mathrm{\mu }$m diameter, matches the sample length, 3 mm. At
higher powers, filaments are observed in the output facet images,
indicating that the collapse occurred earlier in the crystal.
Following the application of a magnetic field, the onset of collapse
at the output facet is observed at \emph{lower} powers [Figs.
4(c-e)], with the largest collapse-acceleration effect recorded at a
magnetic field of $H=400$ G [see Fig. 4(d)], corresponding to half
the saturation field of YIG \cite{isolator}. This observation agrees
with the fact that, when YIG is excited by light traveling
perpendicular to the magnetic field, the linear birefringence is
stronger than its circular counterpart \cite{MO1,CM}. A qualitative
characterization of the output beam polarization state has also
shown that without an external magnetic field the beam remained
linearly polarized, while with an application of the field the beam
became elliptically polarized, with the ellipticity growing with the
field. This indicates the certain presence of magnetically-induced
polarization dynamics in the crystal.

\begin{figure}[t]
\includegraphics[width=8cm]{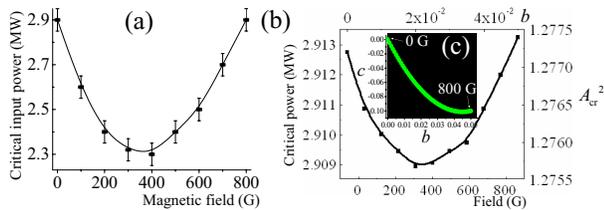}
\caption{ (Color online) (a) Measured critical power required for
the onset of the collapse in the output facet, as a function of the
applied magnetic field. (b) Numerically calculated squared critical
amplitude of the input beam ($A_{\mathrm{cr}}^{2}$) necessary for
the onset of the collapse at $z=7.3$, as a function of the
birefringences corresponding to crystalline YIG. (c) The calculated
dependence between $ b$ and $c$ in YIG, used in (b).}
\end{figure}

Figure 5(a) shows the measured critical power, corresponding to the
data of Fig. 4, that was required for the onset of collapse at the
output facet, as a function of the magnetic field. In Fig. 5(b),
this is compared to results obtained by the numerical solutions of
Eqs. (\ref{eqs}), where for each set of birefringence values
(corresponding to a given magnetic field) we find the critical
amplitude $A_{\mathrm{cr}}$ of the input beam for which the collapse
occurs at the propagation distance corresponding to the output facet
($z=7.3$). Figure 5(c) again displays the relation between the
normalized birefringence parameters of the YIG crystal used, cf. the
dashed curves in Fig. 2. The behavior of the collapse detuning is
similar in Figs. 5(a) and 5(b), even though the theoretical model
did not take into consideration magnetically-induced losses.
Specifically, the Cotton-Mouton effect is always accompanied by
linear magnetic dichroism \cite{CM}, and the Faraday effect entails
circular magnetic dichroism \cite{Faraday-YIG,isolator}, both of
which are weak but present at $\lambda=1.2$ $\mathrm{\mu}$m.

In conclusion, we have investigated the combined effects of circular
and linear birefringences on the propagation of collapsing (2+1)D
beams in self-focusing bulk Kerr media, and have shown that the
onset of collapse can be accelerated, delayed, or suppressed,
depending on the relative birefringence strengths. Experimentally,
we have demonstrated a controlled acceleration of the collapse at
the output facet of a ferrimagnetic YIG crystal, following the
application of an external magnetic field which induces the
birefringences. The direct observation of magnetization-induced
effects in collapsing beams provides a unique demonstration of an
all-optical magnetically-controlled lensing mechanism, pioneering
the use of MO crystals in nonlinear optics experiments. Finally,
since Eqs. (\ref{eqs}) are also Gross-Pitaevskii equations
describing a binary BEC with linear interconversion \cite{Kennedy},
accounted for by the coefficient $c$, similar phenomena may also be
observed in nonlinear matter-wave dynamics.

This research was supported by NSERC and TeraXion (Canada). YL and
KR respectively acknowledge support from FQRNT-MELS and IOF Marie
Curie fellowships.\\$^{*}$ Corresponding author: yoli@emt.inrs.ca.

\end{document}